# Imbalanced spin couplings in the copper hexamer compounds $A_2Cu_3O(SO_4)_3$ ($A_2$=$Na_2$, NaK, $K_2$)

A. Furrer[1], A. Podlesnyak[2], E. Pomjakushina[3], and V. Pomjakushin[1]

[1] Laboratory for Neutron Scattering, Paul Scherrer Institut, CH-5232 Villigen PSI, Switzerland

[2] Neutron Scattering Division, Oak Ridge National Laboratory, Oak Ridge, Tennessee 37831, USA

[3] Laboratory for Multiscale Materials Experiments, Paul Scherrer Institut, CH-5232 Villigen PSI, Switzerland

The minerals $A_2Cu_3O(SO_4)_3$ ($A_2$=$Na_2$, NaK, $K_2$) constitute quantum spin systems with copper hexamers as basic structural units. Strong intra-hexamer spin couplings give rise to an effective triplet ground-state. Weak inter-hexamer spin couplings are responsible for two-dimensional long-range magnetic order in the (b,c)-plane below $3.0<T_c<4.7$ K. We investigated the magnetic excitations at T=1.5 K by inelastic neutron scattering (INS). The INS technique was based on the observation of wavevector-dependent slices in reciprocal space in order to selectively probe the magnetic signals in different Brillouin zones with different weight. Due to the imbalance of the spin couplings, the data analysis relied on a model in which the inter-hexamer spin couplings are treated perturbatively on top of the exact S=1 ground state. The inter-hexamer spin couplings turn out to be ferromagnetic.

Quantum magnetism and related spin-frustration phenomena are an active field in present condensed-matter research, which has been largely inspired by the discovery of naturally occuring minerals [1]. The minerals $A_2Cu_3O(SO_4)_3$ ($A_2$=$Na_2$, NaK, $K_2$) are novel candidates to advance this field. They are of particular interest, since they are composed of molecular-like copper hexamers with strong intra-hexamer and weak inter-hexamer interactions, so that the interplay of different energy scales and dimensionalities can lead to unusual phenomena. Fujihala *et al.* [2] recognized the fascinating magnetic properties of the title compounds for $A_2$=$K_2$, which was proposed to be a one-dimensional antiferromagnet with $Cu_6$ chains along the b-direction involving a triplet ground-state, giving rise to a novel Haldane system [3] with effective spin S=1 at low temperatures. The S=1 triplet ground-state was confirmed by inelastic neutron scattering (INS) experiments performed for the title compounds with $A_2$=$Na_2$ and $A_2$=$K_2$ [4,5], resulting in detailed values of the intra-hexamer exchange parameters $J_{nm}$ marked in Fig. 1 (top) and listed in Table I. The $Cu_6$ chain picture was questioned by Nekrasova *et al.* [6] who proposed two-dimensional antiferromagnetic interactions between the copper hexamers along the diagonals in the (b,c)-plane, thereby challenging Fujihala's description of the title compounds in terms of Haldane spin chains. Both Fujihala's and Nekrasova's conclusions, the latter being supported by density-functional-theory band-structure calculations, are essentially based on magnetic susceptibility, magnetization, and heat capacity data, which constitute integral properties and do not provide reliable information on the magnetic couplings. However, no efforts were made so far to directly determine the inter-hexamer exchange parameters $j_y$ and $j_z$ marked in Fig. 1 (top) which are important to discriminate between one-dimensional and two-dimensional magnetic ordering at low temperatures.

INS studies of the dispersive magnetic excitations (spinwaves) are usually carried out with use of large enough single crystals, which are not available for the title compounds. Here we demonstrate that a detailed parametrization of the spinwaves is possible by using polycrystalline samples. Our method is based on observing wavevector-dependent slices in reciprocal space in order to selectively probe the magnetic signals in different Brillouin zones with different weight. The shapes of the observed energy spectra are found to vary substantially by changing the wavevector-range, thereby reflecting all the details of the magnetic excitations, so that an unambiguous determination of the inter-hexamer exchange parameters $j_y$ and $j_z$ becomes possible.

The compounds $A_2Cu_3O(SO_4)_3$ ($A_2$=$Na_2$, $K_2$, NaK) crystallize in the monoclinic space group C2/c. Details of the synthesis and characterization of the polycrystalline samples are described in Ref. [4]. The copper hexamers form a two-dimensional magnetic network parallel to the (b,c)-plane. The unit cell contains two copper hexamers separated by $\rho$=[(1-2y)b,c/2] as visualized in Fig. 1 (top). The lattice parameters (a,b,c,β) as well as the critical temperatures $T_c$ for magnetic ordering are listed in Table I, and the fractional atomic parameters (x,y,z) are given in Refs. [4,7].

The INS experiments were carried out with use of the high-resolution time-of-flight spectrometer CNCS [8] at the spallation neutron source (SNS) at Oak Ridge National Laboratory. The incoming neutron energy was 3.32 meV, yielding an instrumental energy resolution of about 0.1 meV. Fig. 2(a) shows the temperature dependence of the energy spectra observed for $A_2$=NaK. Below $T_c$=4.7 K there is clear evidence of gapped energy spectra, which extend from 0.8 meV to 2.0 meV. The wavevector dependence of the energy spectra observed for $A_2$=$Na_2$ is shown in Fig. 2(b). The signal above the gap ranges from 0.6 meV to 1.6 meV. Both the shape and the integrated intensities strongly vary upon changing the modulus of the scattering vector **Q**. For $A_2$=$K_2$ the results are identical within experimental error.

The exchange couplings of the title compounds feature a strong imbalance of the relevant exchange parameters $J_{nm}$ and $j_y$, $j_z$, so that the application of the classical spinwave theory turns out to be inadequate. Therefore, our model treats the dominant parameters $J_{nm}$ exactly, providing a triplet ground state with $S=1$ [4,5]. The additional weak inter-hexamer interactions $j_y$ and $j_z$ are considered perturbatively to yield the dispersion relations through the Fourier transform of the couplings between the copper hexamers in the (b,c)-plane:

$$E(\mathbf{q}) = \Delta - 2S[J(\mathbf{q}) \pm |J'(\mathbf{q})|] \tag{1}$$

where $\Delta$ corresponds to the splitting of the ground-state triplet due to the molecular field. The Fourier transforms $J(\mathbf{q})$ and $J'(\mathbf{q})$ of the exchange parameters are given by

$$J(\mathbf{q}) = \sum_{n,m} j_{nm} \exp\{i\mathbf{q}\cdot(\mathbf{R}_n - \mathbf{R}_m)\}, \quad J'(\mathbf{q}) = \sum_{n,k} j_{nk} \exp\{i\mathbf{q}\cdot(\mathbf{R}_n - \mathbf{R}_k)\}, \tag{2}$$

where the summation indices nm and k refer to the copper hexamers in the two sublattices. The parameters $j_{nm}$ and $j_{nk}$ correspond to $j_y$ and $j_z$, respectively, when the exchange interactions are restricted to nearest-neighbor hexamers as visualized in Fig. 1 (top). The site vectors are $\mathbf{R}_n=(b,0)$ and $\mathbf{R}_k=\boldsymbol{\rho}=[(1-2y)b,c/2]$. The spinwave dispersion is split into an acoustic and an optic branch associated with the ± sign in Eq. (1). $J(\mathbf{q})$ is always real, while $J'(\mathbf{q})$ is complex in general. For y value of the Cu$_3$ bond, which is y=0.7475(5) for $A_2$=Na$_2$, y=0.7483(7) for $A_2$=NaK, and y=0.7451(7) for $A_2$=K$_2$ [4,7]. These values are very close to 0.75, thus we set $\boldsymbol{\rho}=[-b/2,c/2]$ and $J'(\mathbf{q})$ becomes real. This is totally justified, since by using the effective y-values of the $A_2$-compounds, we have $|\text{Im}\{J'(\mathbf{q})\}|\ll|\text{Re}\{J'(\mathbf{q})\}|$. In this case the spinwave energies are

$$E(\mathbf{q}) = \Delta - [4j_y\cos(\pi q_y)] \pm 8j_z\cos(\pi q_y/2)\cos(\pi q_z/2)] \tag{3}$$

where **q**=($q_y$,$q_z$) with 0≤$q_y$,$q_z$≤1. Fig. 3 demonstrates the separation into lower-energy acoustic and higher-energy optic spinwave branches. Our model treats the inter-hexamer couplings $j_y$ and $j_z$ in a phenomenological manner. Microscopically, there are several contributions to both $j_y$ and $j_z$, but for $j_y$ the dominant couplings are provided by $Cu_1$-O-S-O-$Cu_2$ superexchange bridges, whereas for $j_z$ the superexchange bridges $Cu_1$-O-S-O-$Cu_1$ and $Cu_2$-O-S-O-$Cu_2$ are relevant, see Fig. 1 (top).

The differential neutron cross-section for spinwaves is given by [9]

$$d^2\sigma/(d\Omega d\omega) \propto S(Q)F^2(\mathbf{Q})[1\pm\cos(\boldsymbol{\tau}\cdot\boldsymbol{\rho})] , \qquad (4)$$

where S(Q) is the structure factor of the copper hexamers [5], F(**Q**) the magnetic form factor, and **τ** a reciprocal lattice vector. It follows from Eq. (4) that the Brillouin zones are separated into acoustic and optic zones as visualized in Fig. 1 (bottom).

The analysis of the experimental data was based on the detailed Q-dependence of the energy spectra shown in Fig. 4. The data for $A_2$=$Na_2$ are very similar to those for $A_2$=$K_2$. The integrated intensities nicely follow the sinusoidal Q-dependence according to the structure factor S(Q) for transitions within the ground-state triplet, see Eq. (4). The variation of the range $Q_1$≤Q≤$Q_2$ strongly changes the coverage C of the acoustic (A) and optic (O) Brillouin zones. From bottom to top in Fig. 4 we have $C_A/C_O$=1.78, 0.72, 1.08, and 1.28, which is directly reflected by the Q-dependence of the intensities associated with acoustic Brillouin zones in the lower-energy part of the spectra.

A rough estimate of the spinwave parameters is obtained from selected points in reciprocal space, which can be used as starting parameters in the least-squares fitting procedures. The energies defined by Eq. (3) for **q**=(0,0) determine the lower and upper limit of the observed energy spectra:

**q**=(0,0):  $E_{acoustic}(q) = \Delta - 4j_y - 8j_z$ ,  $E_{optic}(q) = \Delta - 4j_y + 8j_z$ ,

which essentially fixes the value of $j_z$. Large contributions to the energy spectra are expected around **q** vectors where the derivative $dE(\mathbf{q})/d\mathbf{q}=0$. This is the case for **q**=(0,0) as well as for

**q**=(1,1): $E_{acoustic}(\mathbf{q}) = E_{optic}(\mathbf{q}) = \Delta + 4j_y$ ,

see Fig. 3. The energy spectra observed for $A_2$=NaK are considerably different from $A_2$=$K_2$ and $A_2$=$Na_2$ concerning both the energy range and the spectral shape, the latter being due to the small dispersion of the optic spinwave branches along the (1,0)- and (1,1)-directions which results in enhanced intensities at the upper energy limit.

$\Delta$, $j_y$, and $j_z$ were treated as adjustable parameters in the least-squares fitting analysis of the data displayed in Fig. 4. For each wavevector range $Q_1 \leq Q \leq Q_2$, the spinwave energies were collected in pixels of size 0.002×0.002 Å$^{-2}$ and folded with the instrumental energy resolution. The results are listed in Table I and shown in Figs. 3 and 4. Good agreement between the experimental and calculated energy spectra is obtained with overall standard deviations $\chi^2$=1.9 and $\chi^2$=3.0 for $A_2$=$K_2$ and $A_2$=NaK, respectively. Both exchange parameters $j_y$ and $j_z$ turn out to be ferromagnetic; any attempts to reproduce the observed energy spectra with antiferromagnetic and mixed ferromagnetic/antiferromagnetic exchange parameters failed as demonstrated in detail in Ref. [7].

In principle, there is a self-consistent criterion relating the gap parameter $\Delta$ to the Fourier transform of the exchange parameters:

$\Delta = 2S[J(0)+J'(0)] = 4j_y+8j_z$ . (5)

As shown in Table I, the calculated values for $2S[J(0)+J'(0)]$ are slightly smaller than $\Delta$. The difference is probably due to a gap enhancement D resulting from the single-molecule axial anisotropy, which was suggested to be present for the copper hexamers with D=0.064 meV (for $A_2$=$Na_2$ and $A_2$=$K_2$) and D=0.079 meV (for $A_2$=NaK) [6]. The molecular-field parameter 2S[J(0)+J'(0)] for $A_2$=NaK is larger than for $A_2$=$Na_2$ and $A_2$=$K_2$, which explains the difference of the critical temperatures $T_c$ for long-range magnetic order, see Table I.

The results of the present work are in strong contradiction to earlier findings reported in Refs. [2,6]. Our analysis gives compelling evidence for superexchange inter-hexamer couplings both along the b-axis and along the diagonals in the (b,c)-plane, and the corresponding exchange parameters $j_y$ and $j_z$ are ferromagnetic. This implies that the two-dimensional magnetic structure of the $A_2$-compounds should reflect its basically ferromagnetic nature. Studies along these lines have been presented by Hase *et al*. [10] based on neutron diffraction experiments performed for $A_2$=$K_2$ at low temperatures, indicating the presence of two-dimensional magnetic order with propagation vector **k**=(0,0,0) and ferromagnetic spin alignments along the b-axis.

The compounds $A_2Cu_3O(SO_4)_3$ constitute a rare example of isostructural systems for which the physical properties of the mixed compound $A_2$=NaK cannot be interpolated between the pure compounds $A_2$=$Na_2$ and $A_2$=$K_2$. This is evident, *e.g*., for the lattice parameters b as well as for the magnetic ordering temperatures $T_c$ which for $A_2$=NaK lie outside the limits given by the pure compounds, see Table I. The anomalously low value of b has an effect on the inter-hexamer coupling parameter $j_y$ which turns out to be twice as large compared to $A_2$=$Na_2$ and $A_2$=$K_2$, whereas the coupling parameters $j_z$ are very similar for all $A_2$ compounds.

In conclusion, our work demonstrates that a reliable determination of spin-coupling parameters requires the use of direct methods such as INS spectroscopy. This is particularly true for magnetic compounds with unbalanced and frustrated spin interactions, where *ab initio* calculations have to be considered with caution. We introduced an experimental technique to analyze

INS data taken for polycrystalline samples, which will be a very useful alternative in cases where single-crystals are not available.

TABLE I: Lattice parameters a, b, c, β, and unit cell volume V of the compounds $A_2Cu_3O(SO_4)_3$ determined by neutron diffraction at T=1.5 K (Ref. [4] for $A_2=Na_2$ and $A_2=K_2$, present work for $A_2$=NaK). Critical temperature $T_c$ for magnetic ordering [6]. Intra-hexamer exchange parameters $J_{nm}$ determined by neutron spectroscopy (Ref. [5] for $A_2=Na_2$ and $A_2=K_2$, to be published for $A_2$=NaK). Parameters $\Delta$, $j_y$, and $j_z$ of Eq. (3) determined in the present work. 2S[J(0)+J'(0)] denotes the molecular-field parameter.

|  | $Na_2Cu_3O(SO_4)_3$ | $NaKCu_3O(SO_4)_3$ | $K_2Cu_3O(SO_4)_3$ |
|---|---|---|---|
| a [Å] | 17.21406(67) | 18.4730(8) | 18.97550(66) |
| b [Å] | 9.37286(35) | 9.3644(4) | 9.50038(35) |
| c [Å] | 14.37014(54) | 14.3146(6) | 14.19721(51) |
| β [°] | 111.84364(75) | 113.9642(10) | 110.49150(85) |
| V [Å$^3$] | 2152.61 | 2262.81 | 2375.66 |
| $T_c$ [K] | 3.4 | 4.7 | 3.0 |
| $J_{11}$ [meV] | 1.3(2) | 1.8(2) | 1.6(4) |
| $J_{13a}$ [meV] | -4.7(3) | -4.6(3) | -4.6(3) |
| $J_{13b}$ [meV] | -8.3(3) | -8.1(3) | -7.9(3) |
| $J_{22}$ [meV] | 2.2(2) | 2.9(2) | 2.5(3) |
| $J_{23a}$ [meV] | -5.3(3) | -5.1(3) | -5.2(3) |
| $J_{23b}$ [meV] | -8.3(3) | -8.1(3) | -7.9(3) |
| $J_{33}$ [meV] | 11.5(1.5) | 12.3(1.5) | 12.4(1.5) |
| $\Delta$ [meV] | 1.22(5) | 1.51(3) | 1.24(4) |
| $j_y$ [meV] | 0.025(4) | 0.050(5) | 0.026(3) |
| $j_z$ [meV] | 0.054(4) | 0.062(5) | 0.056(3) |
| 2S[J(0)+J'(0)] [meV] | 1.064 | 1.392 | 1.104 |

**FIGURE CAPTIONS**

FIG. 1. (Color online) Top: Projection of the copper hexamers onto the (y,z)-plane. The dashed and full arrows mark the relevant intra- and inter-hexamer exchange parameters $J_{nm}$ and $j_y$, $j_z$, respectively. Bottom: Projection of the reciprocal lattice of the $A_2$-compounds onto the (b*,c*)-plane. The circles and squares correspond to the centers of the acoustic and optic Brillouin zones shaded in light red and blue colors, respectively. The blue lines define the coverage of the Brillouin zones in neutron scattering experiments for moduli of the scattering vector **Q** in the range $Q_1 \leq Q \leq Q_2$.

FIG. 2. (Color online) (a) Temperature dependence of the energy spectra observed for $NaKCu_3O(SO_4)_3$. The dashed curve corresponds to a power law describing the tail of the elastic line. (b) Wavevector dependence of the energy spectra observed for $Na_2Cu_3O(SO_4)_3$ at T=1.5 K. The dashed curves are as in (a). For clarity, the data of the upper three spectra are enhanced by 2.5, 5, and 7.5 intensity units.

FIG. 3. (Color online) Spinwave dispersions of the $A_2$-compounds along the three symmetry directions in the (b*,c*)-plane.

FIG. 4. (color online) Energy spectra observed for $A_2Cu_3O(SO_4)_3$ with $A_2=K_2$ (a) and $A_2=NaK$ (b). A power-law background is subtracted from the experimental data (see Fig. 2). The lines correspond to least-squares fits based on Eqs. (3) and (4). For clarity, the data of the upper three spectra are enhanced by 3, 6, and 9 intensity units.

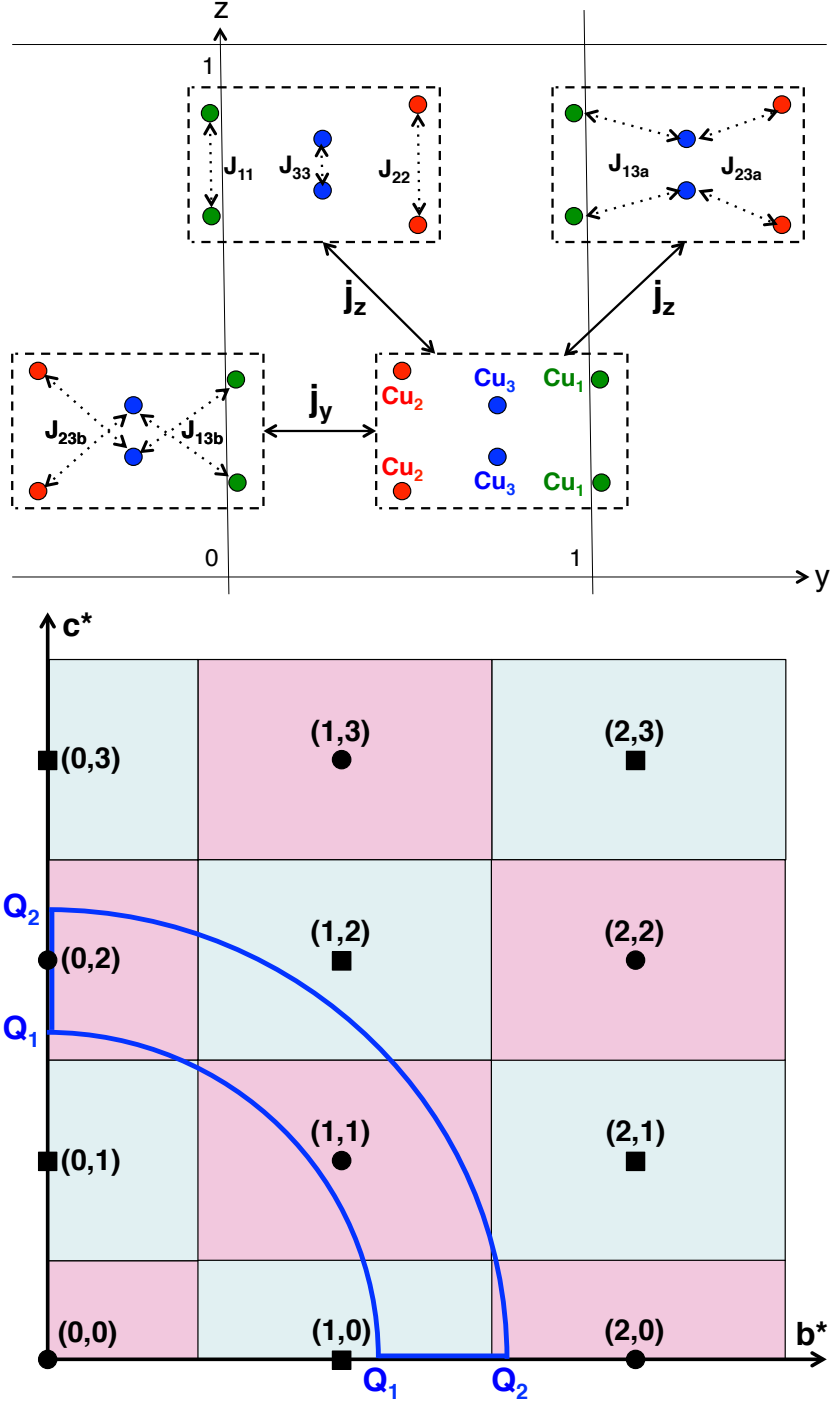

FIG. 1. (Color online) Top: Projection of the copper hexamers onto the (y,z)-plane. The dashed and full arrows mark the relevant intra- and inter-hexamer exchange parameters $J_{nm}$ and $j_y$, $j_z$, respectively. Bottom: Projection of the reciprocal lattice of the $A_2$-compounds onto the ($b^*$,$c^*$)-plane. The circles and squares correspond to the centers of the acoustic and optic Brillouin zones shaded in light red and blue colors, respectively. The blue lines define the coverage of the Brillouin zones in neutron scattering experiments for moduli of the scattering vector **Q** in the range $Q_1 \leq Q \leq Q_2$.

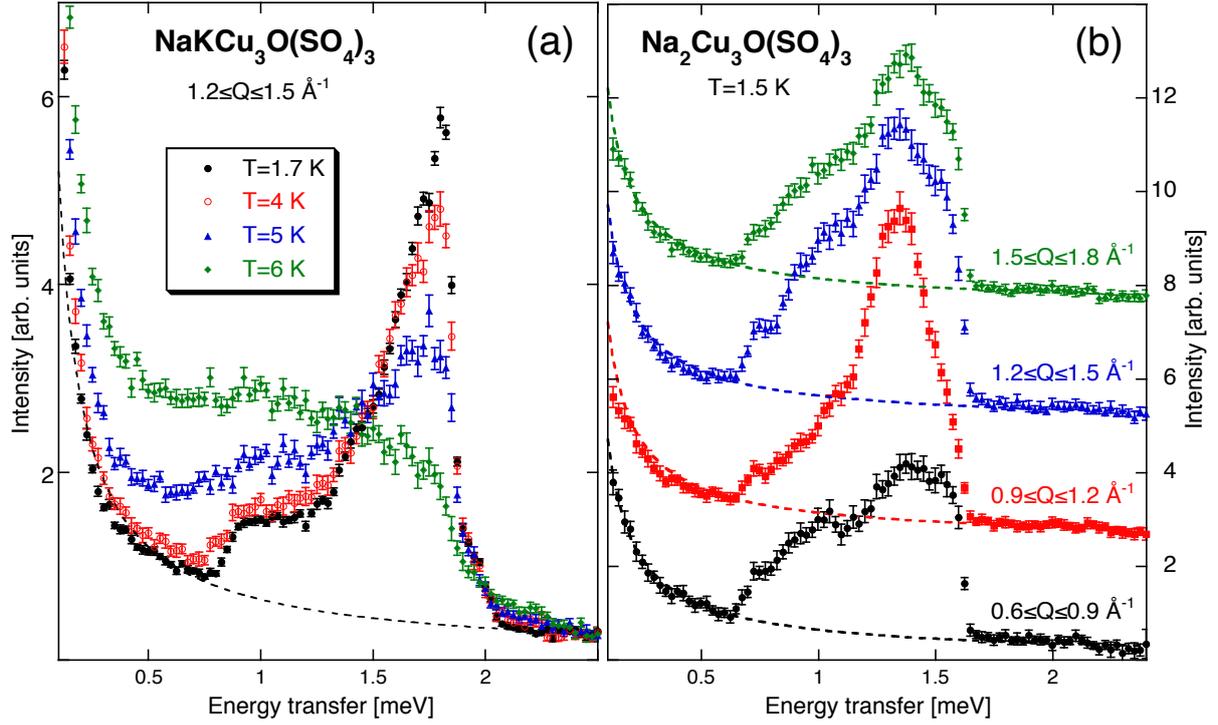

FIG. 2. (Color online) (a) Temperature dependence of the energy spectra observed for $NaKCu_3O(SO_4)_3$. The dashed curve corresponds to a power law describing the tail of the elastic line. (b) Wavevector dependence of the energy spectra observed for $Na_2Cu_3O(SO_4)_3$ at T=1.5 K. The dashed curves are as in (a). For clarity, the data of the upper three spectra are enhanced by 2.5, 5, and 7.5 intensity units.

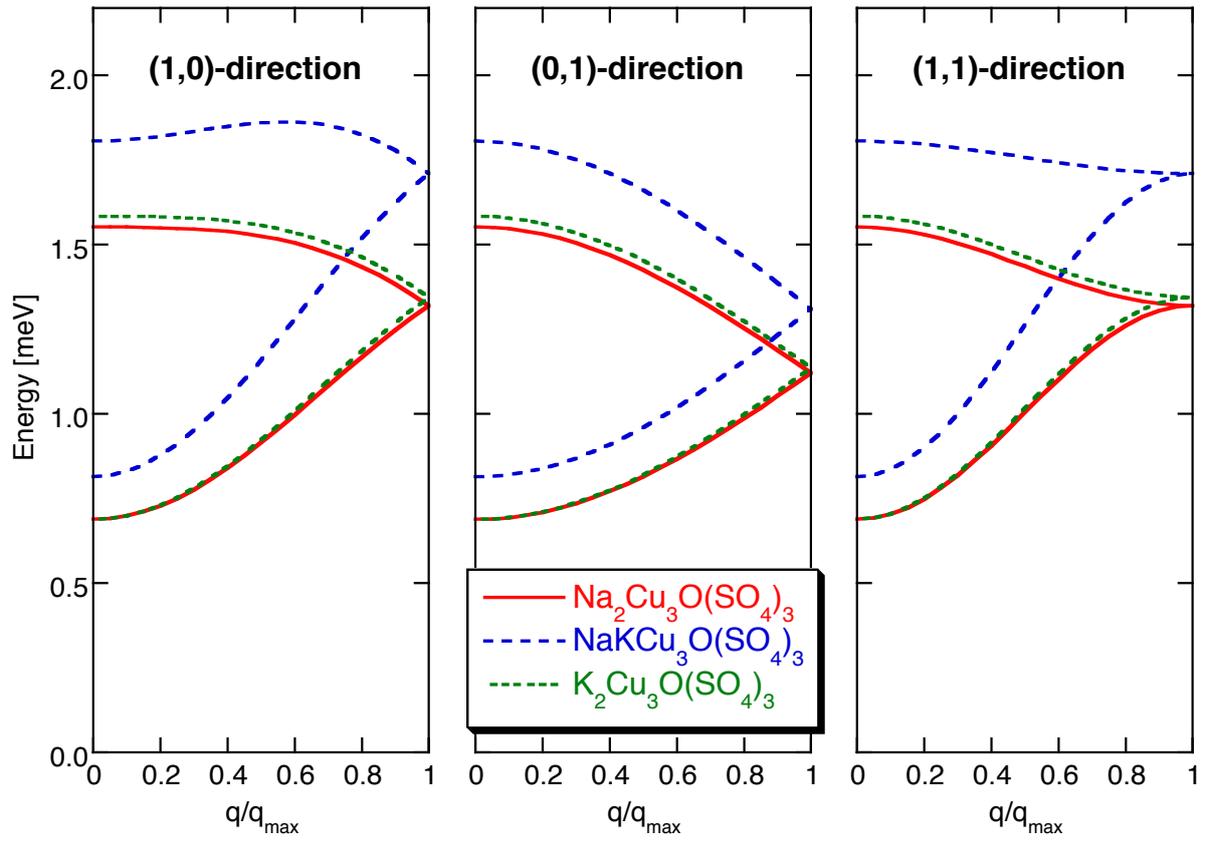

FIG. 3. (Color online) Spinwave dispersions of the $A_2$-compounds along the three symmetry directions in the $(b^*, c^*)$-plane.

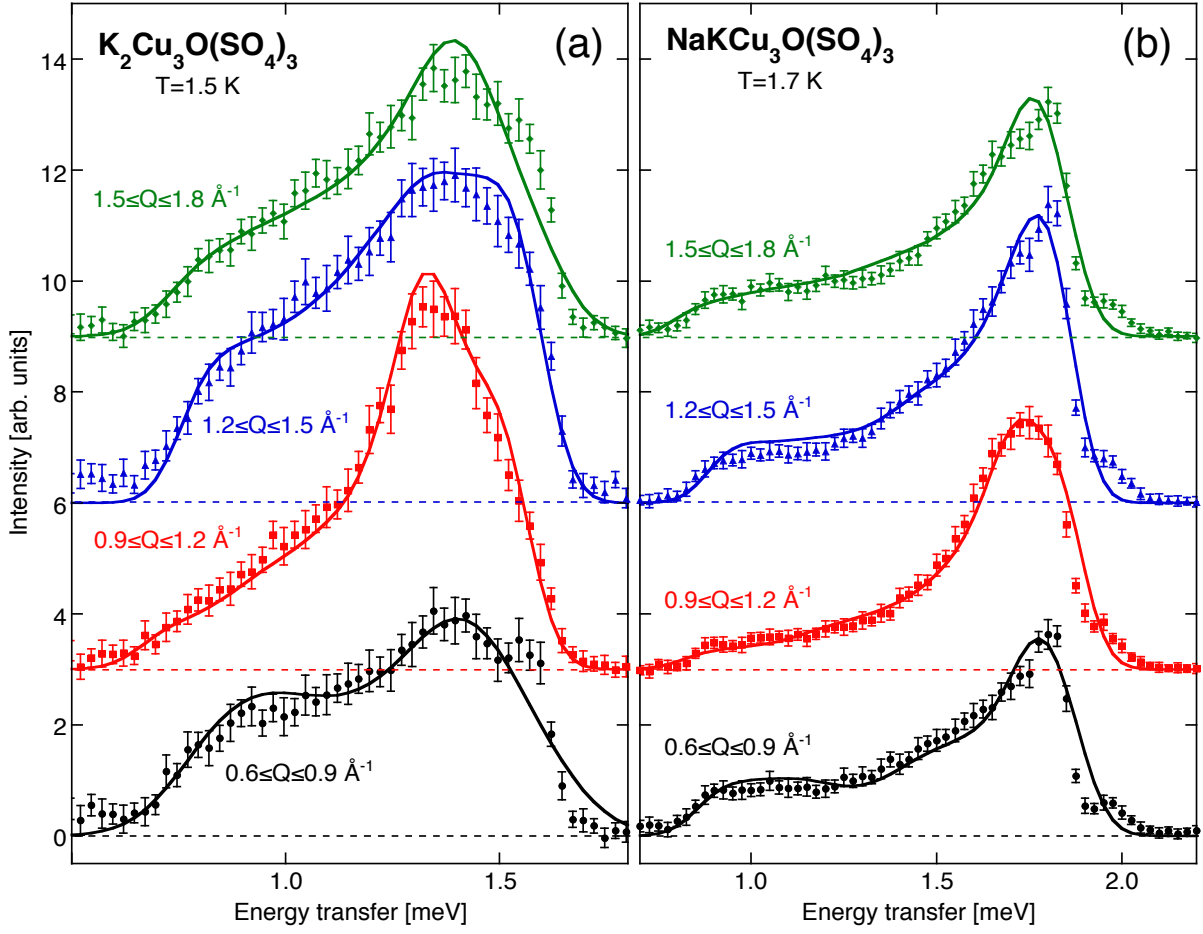

FIG. 4. (color online) Energy spectra observed for $A_2Cu_3O(SO_4)_3$ with $A_2=K_2$ (a) and $A_2=NaK$ (b). A power-law background is subtracted from the experimental data (see Fig. 2). The lines correspond to least-squares fits based on Eqs. (3) and (4). For clarity, the data of the upper three spectra are enhanced by 3, 6, and 9 intensity units.

# Supplemental Material for
# Imbalanced spin couplings in the copper hexamer compounds
# $A_2Cu_3O(SO_4)_3$ ($A_2$=$Na_2$, NaK, $K_2$)

A. Furrer[1], A. Podlesnyak[2], E. Pomjakushina[3], and V. Pomjakushin[1]

[1] Laboratory for Neutron Scattering, Paul Scherrer Institut, CH-5232 Villigen PSI, Switzerland
[2] Neutron Scattering Division, Oak Ridge National Laboratory, Oak Ridge, Tennessee 37831, USA
[3] Laboratory for Multiscale Materials Experiments, Paul Scherrer Institut, CH-5232 Villigen PSI, Switzerland

## 1. NEUTRON DIFFRACTION

Neutron powder diffraction experiments were performed for $NaKCu_3O(SO_4)_3$ with use of the high-resolution diffractometer HRPT at the spallation neutron source SINQ at PSI Villigen with neutron wavelength $\lambda$=2.45 Å. The refinements of the crystal structures were carried out with the program FULLPROF. The resulting lattice parameters are listed in Tables S1 and S2. For comparison, Table S1 also contains earlier data obtained for $A_2Cu_3O(SO_4)_3$ (A=Na, K), see Ref. [4] of the main paper.

## 2. REMARKS ON THE DATA ANALYSIS

The fitting of the energy spectra was based on Eqs. (3) and (4) given in the main paper. The adjustable parameters are $\Delta$, $j_y$, and $j_z$, which are related through the self-consistent criterion Eq. (5). In principle, both $j_y$ and $j_z$ may be ferromagnetic or antiferromagnetic. Below we demonstrate that antiferromagnetic interactions can be excluded.

The total width of the energy spectra observed for the compounds $A_2Cu_3O(SO_4)_3$ (A=Na, K) is about 1 meV which is determined by the minimal and maximal energies of the spinwaves. This fixes the value of $j_z$ to $0.05 \leq |j_z| \leq 0.06$ meV, and the value of $j_y$ is fixed to $0.02 \leq |j_y| \leq 0.03$ meV in order to fulfill the self-consistent criterion. Similar considerations apply for the compound $NaKCu_3O(SO_4)_3$. Fig. S1 shows the calculated energy spectra ($0.9 \leq Q \leq 1.2$ Å$^{-1}$) for ferromagnetic (e), antiferromagnetic (h), and mixed ferromagnetic/antiferromagnetic (f,g) exchange parameters, together with the corresponding spinwave dispersions (a-d). Excellent agreement with the experimental data is only obtained for ferromagnetic interactions (e) with a standard deviation $\chi^2$=1.0. The results of the calculations (g) and (h) can readily be discarded. The calculated shape (f) has some similarity with the experimental data, but there is a considerable disagreement with the observed intensity distribution due to the interchange of acoustic and optic Brillouin zones, so that the calculation (f) has to be discarded as well. In conclusion, there is compelling evidence for the ferromagnetic nature of the exchange parameters $j_y$ and $j_z$.

Fujihala *et al.* (Ref. [2] of the main paper) suggested antiferromagnetic inter-hexamer couplings along the b-axis with $j_y$=-1.55 meV ($j_z$=0), whereas Nekrasova *et al.* (Ref. [6] of the main paper) proposed antiferromagnetic inter-hexamer couplings along the diagonals in the (b,c)-plane with $j_z$=-0.107 meV ($j_y$=0). These parameter sets produce results which totally disagree with both the shape and the width of the energy spectra observed in the present work as shown in Fig. S2.

TABLE S1: Lattice parameters a, b, c, β and unit cell volume V of the compounds $A_2Cu_3O(SO_4)_3$ determined by neutron diffraction (Ref. [4] for $A_2$=$Na_2$ and $A_2$=$K_2$, present work for $A_2$=NaK).

|  | $Na_2Cu_3O(SO_4)_3$ | | $NaKCu_3O(SO_4)_3$ | | $K_2Cu_3O(SO_4)_3$ | |
|---|---|---|---|---|---|---|
|  | T=293 K | T=2 K | T=293 K | T=1 K | T=293 K | T=2 K |
| a [Å] | 17.31714(48) | 17.21406(67) | 18.5522(8) | 18.4730(8) | 19.08335(58) | 18.97550(66) |
| b [Å] | 9.39965(25) | 9.37286(35) | 9.3920(4) | 9.3644(4) | 9.52801(32) | 9.50038(35) |
| c [Å] | 14.39339(39) | 14.37014(54) | 14.3373(6) | 14.3146(6) | 14.20051(44) | 14.19721(51) |
| β [°] | 111.95650(60) | 111.84364(75) | 113.9396(11) | 113.9642(10) | 110.60856(82) | 110.49150(85) |
| V [Å³] | 2172.95 | 2152.61 | 2283.30 | 2262.81 | 2416.80 | 2375.66 |

TABLE S2: Fractional atomic coordinates x, y, z and isotropic displacement factors B of the compound $NaKCu_3O(SO_4)_3$ determined by neutron diffraction at T=1 K. $R_n$ and $\chi^2$ denote the reliability factors.

| Atom | x | y | z | B[Å²] |
|---|---|---|---|---|
| Cu1 | 0.4827(3) | 0.0205(6) | 0.3430(4) | 0.65(7) |
| Cu2 | 0.4834(3) | 0.4724(5) | 0.1381(4) | 0.65(7) |
| Cu3 | 0.4166(3) | 0.7483(7) | 0.2043(3) | 0.65(7) |
| S1 | 0.4957(8) | 0.7514(16) | 0.4873(10) | 0.61(18) |
| S2 | 0.6567(8) | 0.0345(15) | 0.3771(10) | 0.61(18) |
| S3 | 0.3486(8) | 0.4431(14) | 0.2126(9) | 0.61(18) |
| K1 | 0.3245(6) | 0.7595(12) | 0.4573(8) | 1.04(18) |
| Na1 | 0.1938(6) | 0.7399(13) | 0.1272(7) | 1.04(18) |
| O1 | 0.50000 | 0.8907(11) | 0.25000 | 0.67(5) |
| O2 | 0.4382(4) | 0.8272(7) | 0.3963(5) | 0.67(5) |
| O3 | 0.5577(4) | 0.6727(6) | 0.4592(6) | 0.67(5) |
| O4 | 0.4546(3) | 0.6515(7) | 0.5186(5) | 0.67(5) |
| O5 | 0.5961(4) | 0.0596(6) | 0.4144(5) | 0.67(5) |
| O6 | 0.4011(4) | 0.4360(6) | 0.3144(5) | 0.67(5) |
| O7 | 0.3314(4) | 0.6087(8) | 0.1864(5) | 0.67(5) |
| O8 | 0.2718(3) | 0.3869(7) | 0.1872(5) | 0.67(5) |
| O9 | 0.5413(4) | 0.8558(7) | 0.5656(5) | 0.67(5) |
| O10 | 0.6228(4) | 0.0826(7) | 0.2608(5) | 0.67(5) |
| O11 | 0.3807(4) | 0.3986(6) | 0.1372(5) | 0.67(5) |
| O12 | 0.50000 | 0.5992(10) | 0.25000 | 0.67(5) |
| O13 | 0.6698(3) | 0.8755(8) | 0.3727(5) | 0.67(5) |
| O14 | 0.7285(4) | 0.1005(7) | 0.4306(5) | 0.67(5) |

$R_p$=3.42%, $R_{wp}$=4.39%, $R_{exp}$=3.16%, $\chi^2$=1.93

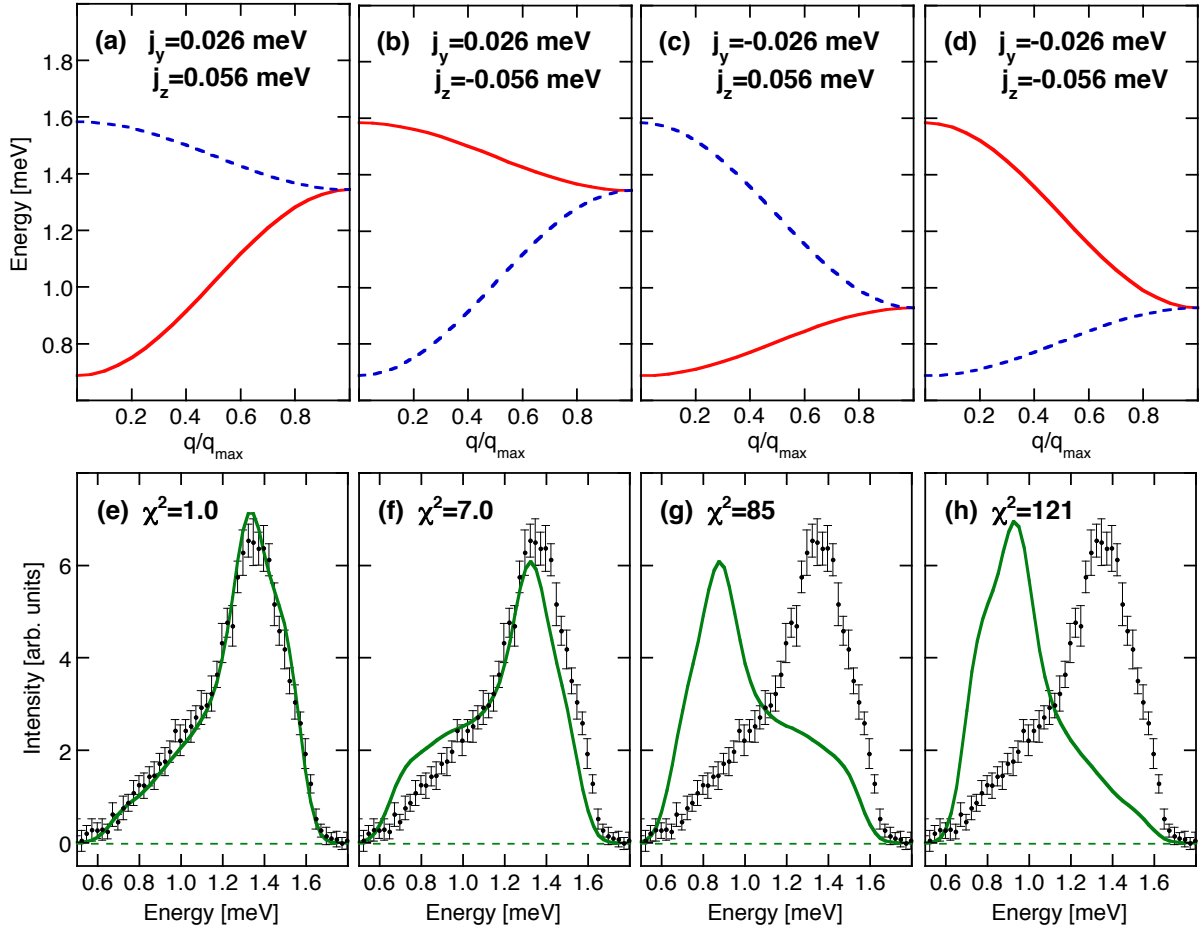

FIG S1. (a-d) Spinwave dispersions of the compound $K_2Cu_3O(SO_4)_3$ along the (1,1)-direction in the (b*,c*)-plane calculated for ferromagnetic and antiferromagnetic exchange parameters $j_y$ and $j_z$. The full and broken lines refer to spinwave branches which can be observed in the acoustic and optic Brillouin zones, respectively. The gap parameters $\Delta$ are adjusted to be in agreement with the lower and upper limits of the observed energy spectra: $\Delta=1.24$ meV for (a) and (b), $\Delta=1.03$ meV for (c) and (d). (e-h) Calculated and observed energy spectra for $0.9 \leq Q \leq 1.2$ Å$^{-1}$. The lines correspond to the calculations for the exchange parameters $j_y$ and $j_z$ given in (a-d). $\chi^2$ denotes the standard deviation.

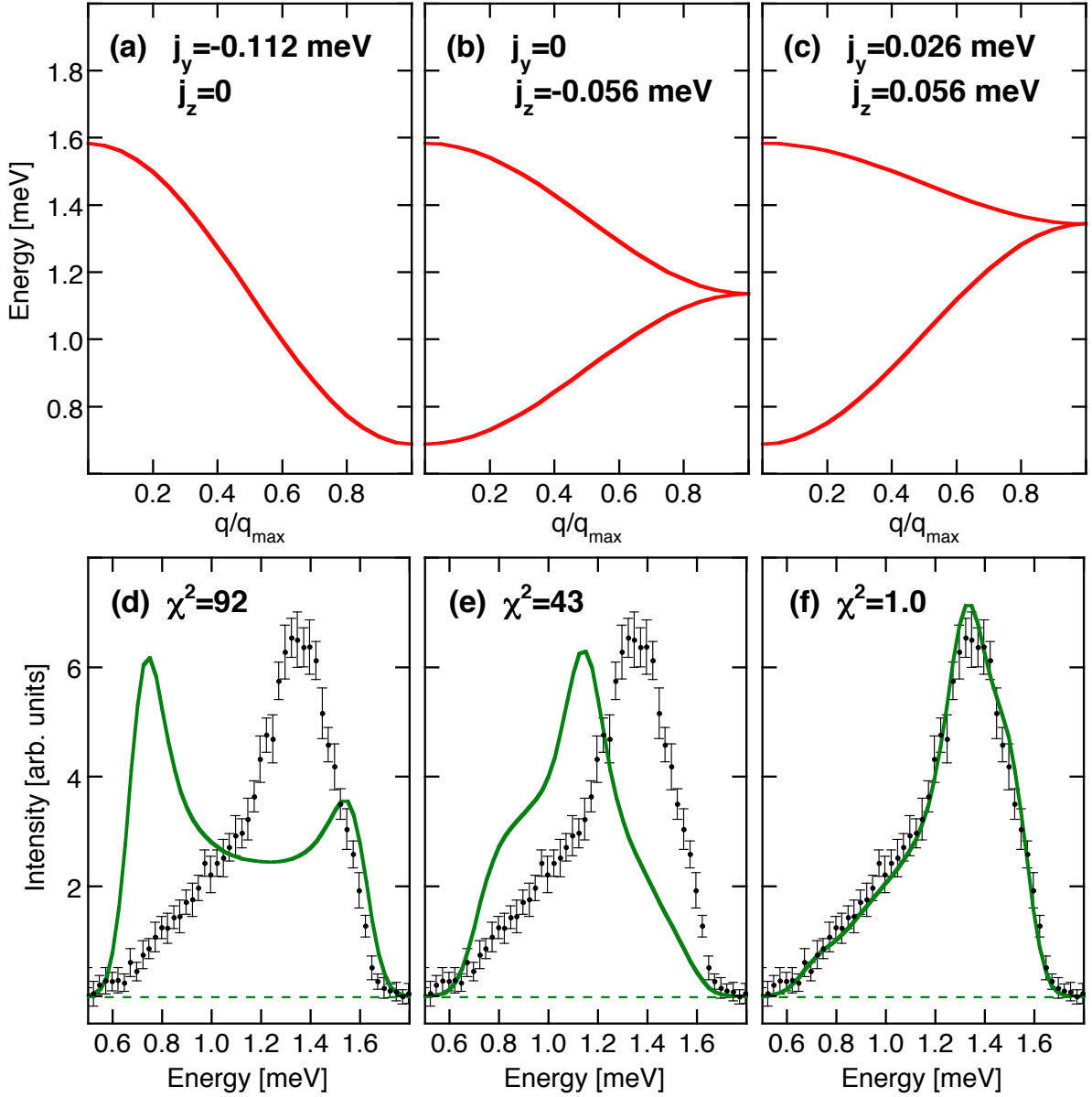

FIG. S2. (a-c) Spinwave dispersions of the $A_2$-compounds along the (1,1) direction in the ($b^*,c^*$)-plane: (a) Model proposed by Fujihala *et al.* (Ref. [2] of the main paper), (b) model proposed by Nekrasova *et al.* (Ref. [6] of the main paper), (c) parameters determined in the present work for $K_2Cu_3O(SO_4)_3$. The parameters $j_y$ and $j_z$ used in (a) and (b) are scaled down from the values given in Refs. [2,6] in order to get agreement with the width of the observed energy spectra. (d-f) Calculated and observed energy spectra for $0.9 \leq Q \leq 1.2$ Å$^{-1}$. The lines correspond to the calculations for the exchange parameters $j_y$ and $j_z$ given in (a-c). $\chi^2$ denotes the standard deviation.